\documentclass[]{elsarticle}

\pdfoutput=1

\usepackage{graphicx}
\usepackage{hyperref}

\newcommand{\fig}[1]{Fig. \ref{#1}}

\begin{document}

\title{Beam tests of a large-scale TORCH time-of-flight demonstrator}

\author[1]{T. H. Hancock\corref{cor1}}\ead{thomas.hancock@physics.ox.ac.uk}
\author[2,5]{\small S. Bhasin}
\author[3]{\small T. Blake}
\author[5]{\small N. Brook}
\author[8]{\small T. Conneely}
\author[2]{\small D. Cussans}
\author[6]{\small C. Frei}
\author[6]{\small R. Forty}
\author[4]{\small E. P. M. Gabriel}
\author[1]{\small R. Gao}
\author[3]{\small T. Gershon}
\author[6]{\small T. Gys}
\author[1]{\small T. Hadavizadeh}
\author[1]{\small N. Harnew}
\author[3]{\small M. Kreps}
\author[8]{\small J. Milnes}
\author[6]{\small D. Piedigrossi}
\author[2]{\small J. Rademacker}
\author[7]{\small M. van Dijk}

\address[1]{\footnotesize Denys Wilkinson Laboratory, University of Oxford, Keble Road, Oxford, OX1 3RH, United Kingdom}
\address[2]{\footnotesize H.H. Wills Physics Laboratory, University of Bristol, Tyndall Avenue, Bristol, BS8 1TL, United Kingdom}
\address[3]{\footnotesize Department of Physics, University of Warwick, Coventry, CV4 7AL, United Kingdom}
\address[4]{\footnotesize School of Physics and Astronomy, University of Edinburgh, James Clerk Maxwell Building, Edinburgh, EH9 3FD, United Kingdom}
\address[5]{\footnotesize University of Bath, Claverton Down, Bath, BA2 7AY, United Kingdom}
\address[6]{\footnotesize CERN, EP Department, CH-1211 Geneva 23, Switzerland}
\address[7]{\footnotesize CERN, EN Department, CH-1211 Geneva 23, Switzerland}
\address[8]{\footnotesize Photek Ltd., 26 Castleham Road, St Leonards on Sea, East Sussex, TN389 NS, United Kingdom}

\cortext[cor1]{Corresponding author}

\begin{abstract}
The TORCH time-of-flight detector is designed to provide particle identification in the momentum range $2-10\,\rm{GeV/c}$ over large areas. The detector exploits prompt Cherenkov light produced by charged particles traversing a $10\,\rm{mm}$ thick quartz plate. The photons propagate via total internal reflection and are focused onto a detector plane comprising position-sensitive Micro-Channel Plate Photo-Multiplier Tubes (MCP-PMT) detectors. The goal is to achieve a single-photon timing resolution of $70\,\rm{ps}$, giving a timing precision of $15\,\rm{ps}$ per charged particle by combining the information from around 30 detected photons. The MCP-PMT detectors have been developed with a commercial partner (Photek Ltd, UK), leading to the delivery of a square tube of active area $53 \times 53\,\rm{mm^2}$ with a granularity of $8 \times 128\,\rm{pixels}$ equivalent. A large-scale demonstrator of TORCH, having a quartz plate of dimensions $660 \times 1250 \times 10\,\rm{mm^3}$ and read out by a pair of MCP-PMTs with custom readout electronics, has been verified in a test beam campaign at the CERN PS. Preliminary results indicate that the required performance is close to being achieved. The anticipated performance of a full-scale TORCH detector at the LHCb experiment is presented.
\end{abstract}

\maketitle

%
\section{Introduction}
\label{sec:Introduction}
\begin{figure}[ht]
\centering
\includegraphics[width=\linewidth]{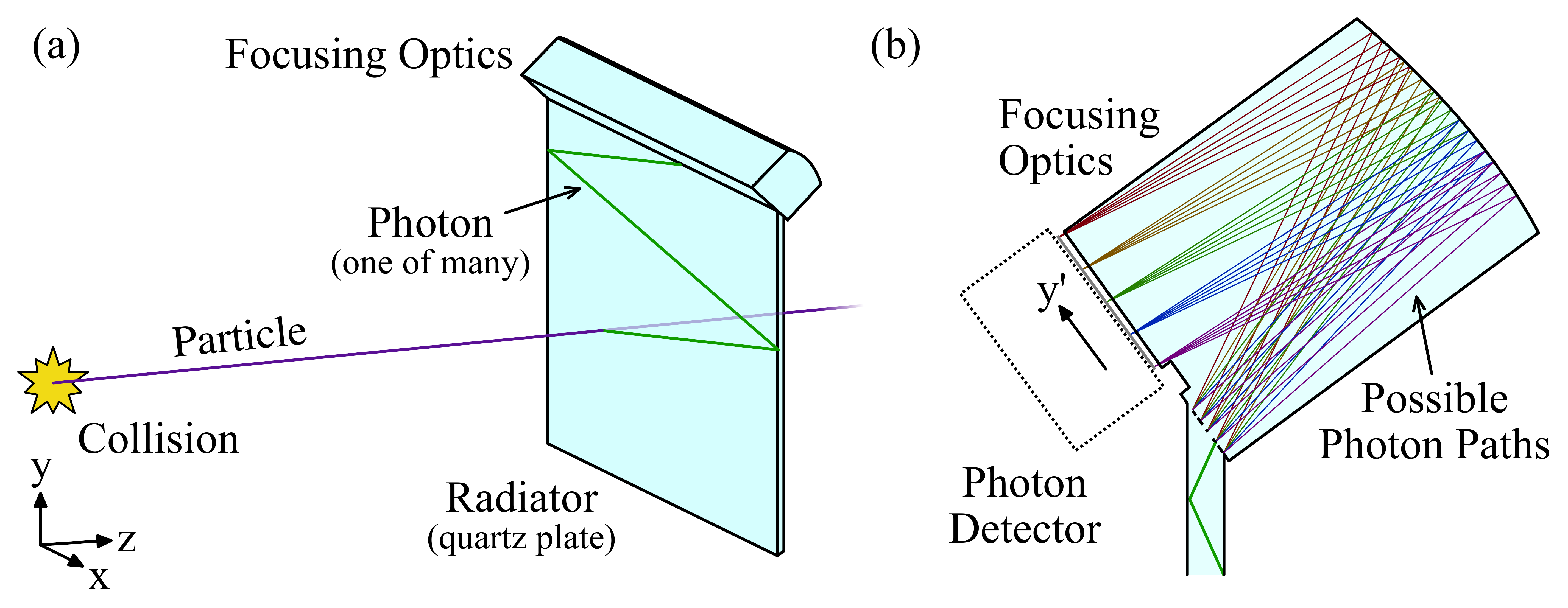}
\caption{Schematics demonstrating the TORCH principle. (a) Particles passing through the radiator plate produce Cherenkov photons which are trapped by total internal reflection and travel to the focusing optics. (b) The focusing optics map photon angle to $y^{\prime}$-position on the photon detector plane, allowing the Cherenkov angle to be determined.}
\label{fig:Detector_Design_Big}
\end{figure}

TORCH is a Time of Flight (ToF) detector which aims to improve low momentum ($2-10\,\rm{GeV/c}$) Particle Identification (PID) for the LHCb experiment \cite{Charles_2011, Brook_2018}. The TORCH detector consists of an array of quartz radiator plate optically coupled to focusing optics at one end, shown in \fig{fig:Detector_Design_Big}. When a charged particle passes through the $10\,\rm{mm}$ thick radiator plate, Cherenkov photons are generated and subsequently trapped by total internal reflection, some travelling to the top of the plate. Here, the focusing optics map the photon angle in the y-z plane to $\rm{y}^{\prime}$ position on the detector, with $\rm{y}^{\prime}$ rotated $36^{\circ}$ from the y-axis (see \fig{fig:Detector_Design_Big}).
Combined with tracking information, this mapping allows the Cherenkov angles to be determined. The position and time of arrival of the Cherenkov photons are measured by custom-designed Micro-Channel Plate Photo-Multiplier Tubes (MCP-PMTs) \cite{Conneely_2015} developed by Photek Ltd (UK). These are square tubes of active area $53 \times 53\,\rm{mm^2}$ with a granularity of $8 \times 128\,\rm{pixels}$-equivalent\footnote{The actual pixelization is $8 \times 64$, but charge-sharing is used to achieve finer resolution.} in x and $\rm{y}^{\prime}$ respectively, read out by electronics employing the NINO and HPTDC chipsets \cite{Akindinov_2004, Anghinolfi_2004}. Combining information from all detected photons, the arrival time of the incident particle can be found, and the photon time of propagation and  particle time of flight determined.

TORCH is proposed for the Upgrades Ib and II of the LHCb detector and will consist of eighteen $660 \times 2500 \times 10\,\rm{mm^3}$ modules positioned approximately $9.5\,\rm{m}$ from the interaction region. Over this distance, the difference in time of flight between pions and kaons at $10\,\rm{GeV/c}$ is $\sim35\,\rm{ps}$, requiring a track time-resolution of $15\,\rm{ps}$ for their clean separation. Expecting 30 detected photons per particle, this requires a single-photon timing resolution of $70\,\rm{ps}$.

%
\section{The Test Beam Campaign}
\label{sec:Beam_Test_Campaign}
In October 2018, a full-width, half-height ($660 \times 1250 \times 10\,\rm{mm^3}$) prototype module (named Proto-TORCH) was demonstrated at the CERN PS East Hall T9 facility, with the aim of measuring the single-photon timing resolution and photon counting efficiency. 
Whilst Proto-TORCH has provision for 11 MCP-PMT detectors, only 2 were instrumented in the first phase of tests described here, located on the left-hand edge when viewing Proto-TORCH from its front face.
Proto-TORCH was mounted on a table, moveable with respect to the beam, and data were taken with the beam incident at various positions on the radiator plate. For all positions, the radiator plate was tilted back by $5\,\rm{^\circ}$ from the vertical to put the photon pattern at a more favourable place on the detector.

\begin{figure}[ht]
\centering
\includegraphics[width=\linewidth]{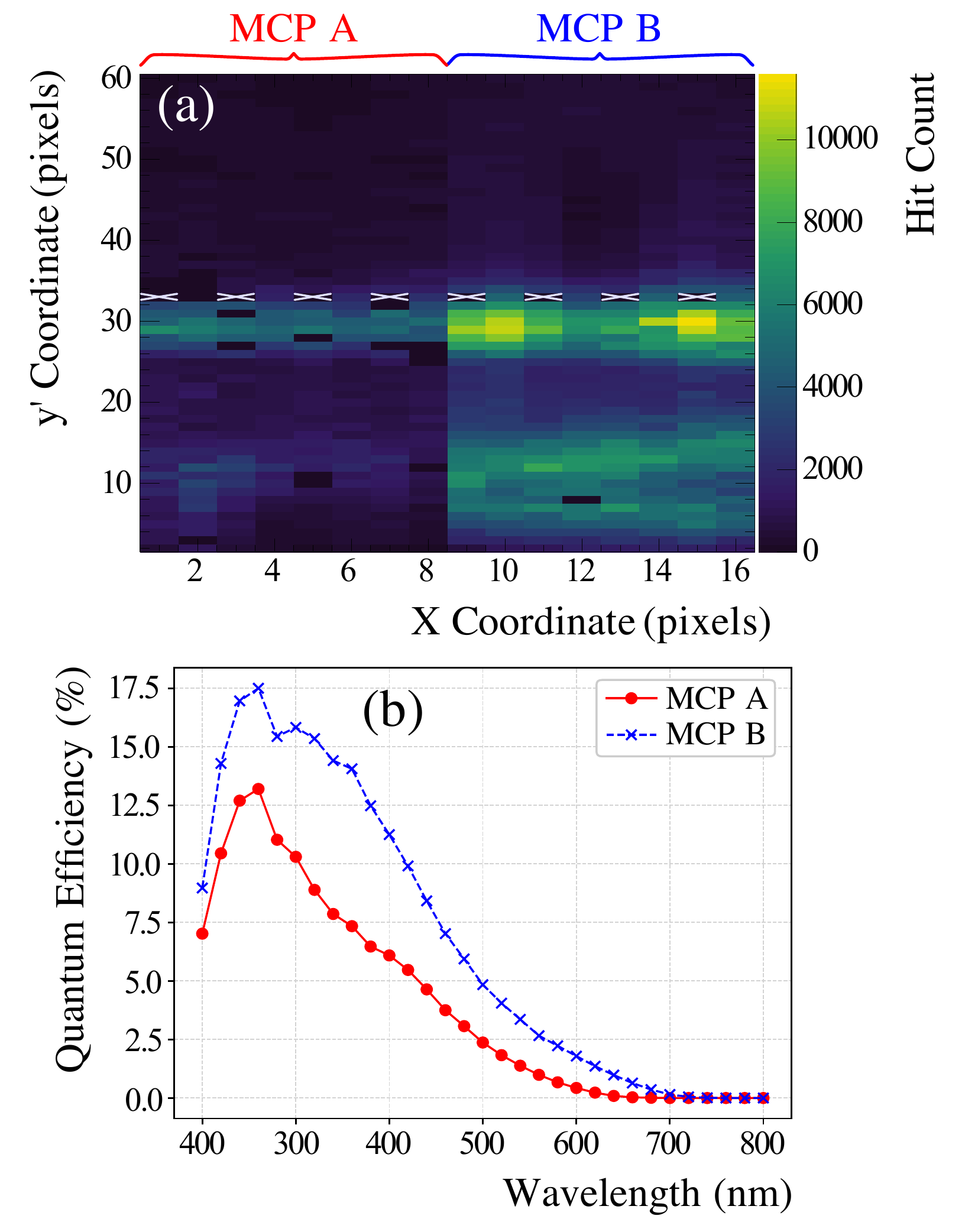}
\caption{
(a) A hit-map recorded using Proto-TORCH with the beam striking the radiator plate at the horizontal mid-point near the bottom. Bands spanning the width of the MCP-PMTs' active area can be seen. The top band spanning the centre of the plot corresponds to light which has travelled from the beam to the MCP-PMT plane without striking the side edges of the radiator plate, while the lower bands correspond to photons reflected off one or more of the side edges. Pixels marked with a cross are deliberately removed to allow the injection of signals from a pair of time reference stations. (b) The measured quantum efficiencies of the two MCP-PMTs, as a function of photon wavelength.
}
\label{fig:Hitmap}
\end{figure}

\fig{fig:Hitmap}a shows a raw hit-map recorded with the beam striking the radiator plate at its horizontal mid-point, $135\,\rm{mm}$ above the bottom face. 
The count rate is lower on the left-hand MCP-PMT due to it having a lower quantum efficiency than the second, as shown in \fig{fig:Hitmap}b.
Empty channels are attributed to NINO chip wire bonds which are unconnected, or to contact issues in the MCP-PMT. Those channels marked with a cross are deliberately removed to allow the injection of signals from a pair of time reference stations.

The two time reference stations are each constructed from thin borosilicate bars, positioned approximately $11\,\rm{m}$ apart. Each produces Cherenkov light which is detected by a single-channel MCP-PMT. In addition to providing a start time, $t_{0}$, for Proto-TORCH, they also provide an independent time of flight measurement which can be used for PID of the beam, composed of approximately 46:54 pions:protons at a momentum of $8\,\rm{GeV/c}$. The beam infrastructure also includes a pair of Cherenkov counters, which provide an alternative source of PID, and a beam telescope to measure the beam profile.

%
\section{Time Resolution}
\label{sec:Time_Resolution}
\begin{figure}[ht]
\centering
\includegraphics[width=0.85\linewidth]{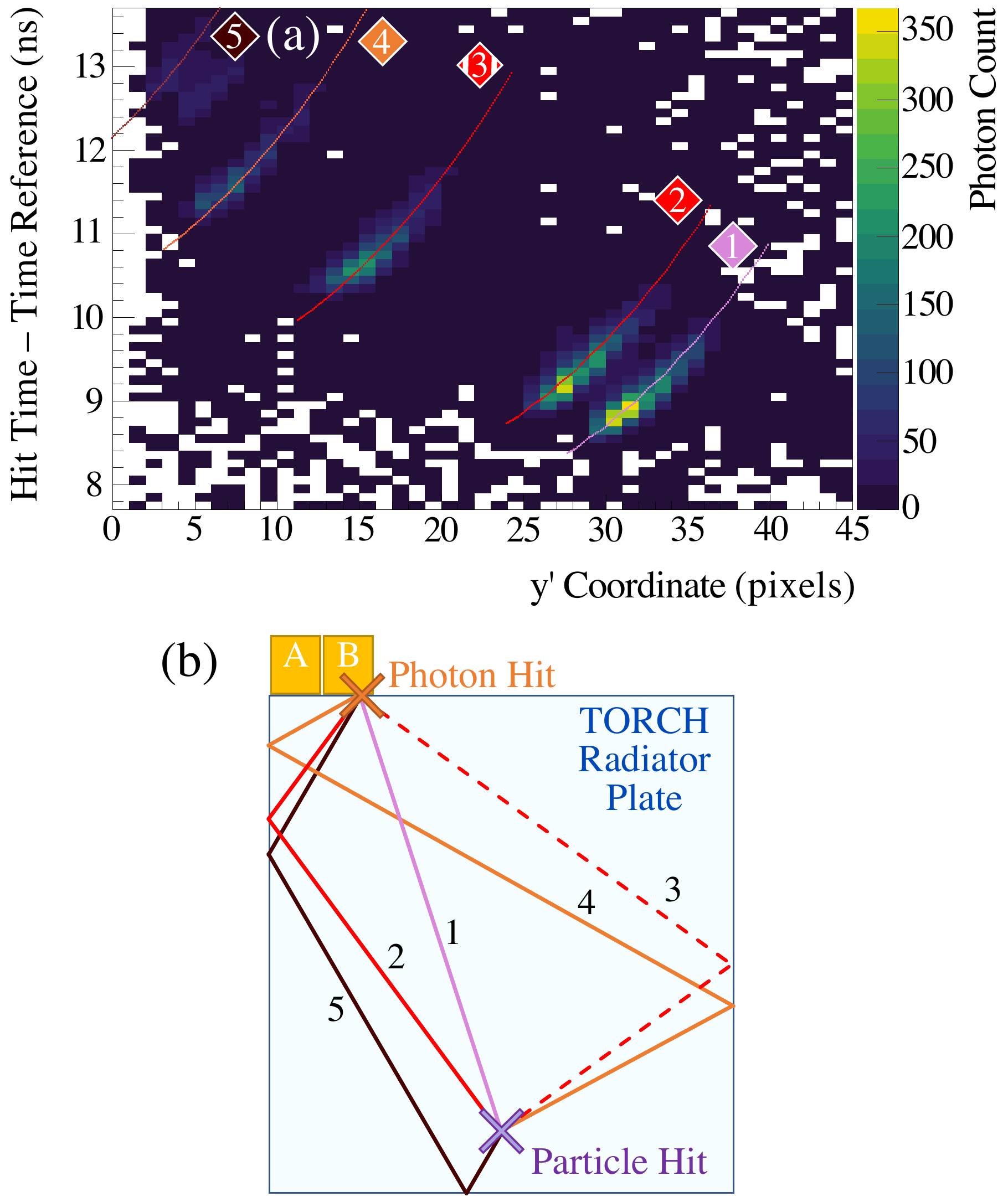}
\caption{
(a) The distribution of photon arrival times as a function of fine-grain pixel number (y$^{\prime}$-position), shown for incident pions detected on column 16 (see \fig{fig:Hitmap}).
The overlaid lines show the predicted times of arrival.
Distinct bands can be seen, corresponding to the photon paths shown in (b).
}
\label{fig:Time_Projection}
\end{figure}

\begin{figure}[ht]
\centering
\includegraphics[width=0.85\linewidth]{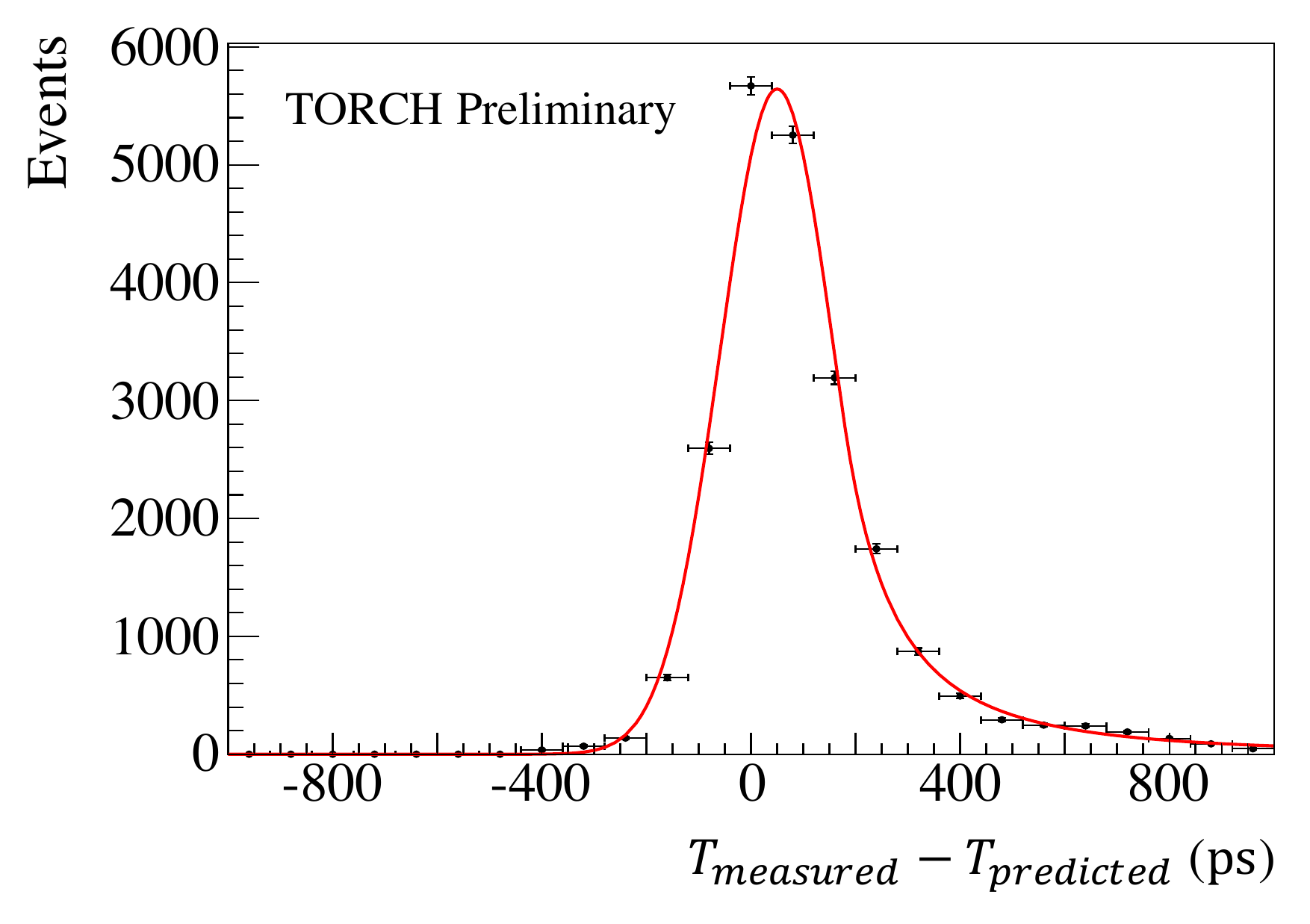}
\caption{
An example residual distribution for incident protons, displaying the predicted time of arrival subtracted from that measured. 
This is shown for photons with no side reflections and detected on column 16 (see \fig{fig:Hitmap}).
A ``Crystal-Ball'' shape (a Gaussian spliced with an exponential tail) is fitted to the data. The width of the Gaussian component is taken as the time resolution of the measurement.
}
\label{fig:Time_Residual}
\end{figure}

The single-photon time resolution of Proto-TORCH is determined by comparing the detected time of arrival for a photon with that predicted from reconstruction. 
\fig{fig:Time_Projection}a shows an example of a scatter plot of the arrival times of  photons as a function of the hit pixel number in the $\rm{y}^{\prime}$ direction for a single MCP-PMT column. 
The results presented here are preliminary.
The overlaid lines show predictions for the various orders of photon reflections, calculated from the incident beam position and momentum, and match the photon paths shown in \fig{fig:Time_Projection}b.
The difference of measured and predicted arrival times results in a residual distribution, an example of which is shown in \fig{fig:Time_Residual}. 
The tail of the distribution is attributed to backscattering of primary photoelectrons in the MCP-PMT and imperfectly calibrated electronics channels. The width is extracted through a fit, giving the time resolution of the measurement, $\sigma_{measured}$. The intrinsic single-photon time resolution of TORCH, $\sigma_{TORCH}$, is then given by
\begin{equation}
\label{eqn:Spread_Contributions}
\sigma_{TORCH}^{2} = \sigma_{measured}^{2} - \sigma_{timeref}^{2} - \sigma_{beam}^{2},
\end{equation}
where $\sigma_{timeref}$ is the resolution of the $t_{0}$ time reference against which Proto-TORCH time-stamps are measured, and $\sigma_{beam}$ is a contribution due to the spread of the beam. This beam correction term is necessary because the time-of-arrival prediction assumes particles enter the radiator plate at a single position. Determined by combining measurements of the beam profile and simulation, and depending on the position of the beam on the radiator plate, the value of $\sigma_{beam}$ ranges from $12.1\pm0.3\,\rm{ps}$ to $32.5\pm0.9\,\rm{ps}$.
 
\begin{figure}[ht]
\centering
\includegraphics[width=0.85\linewidth]{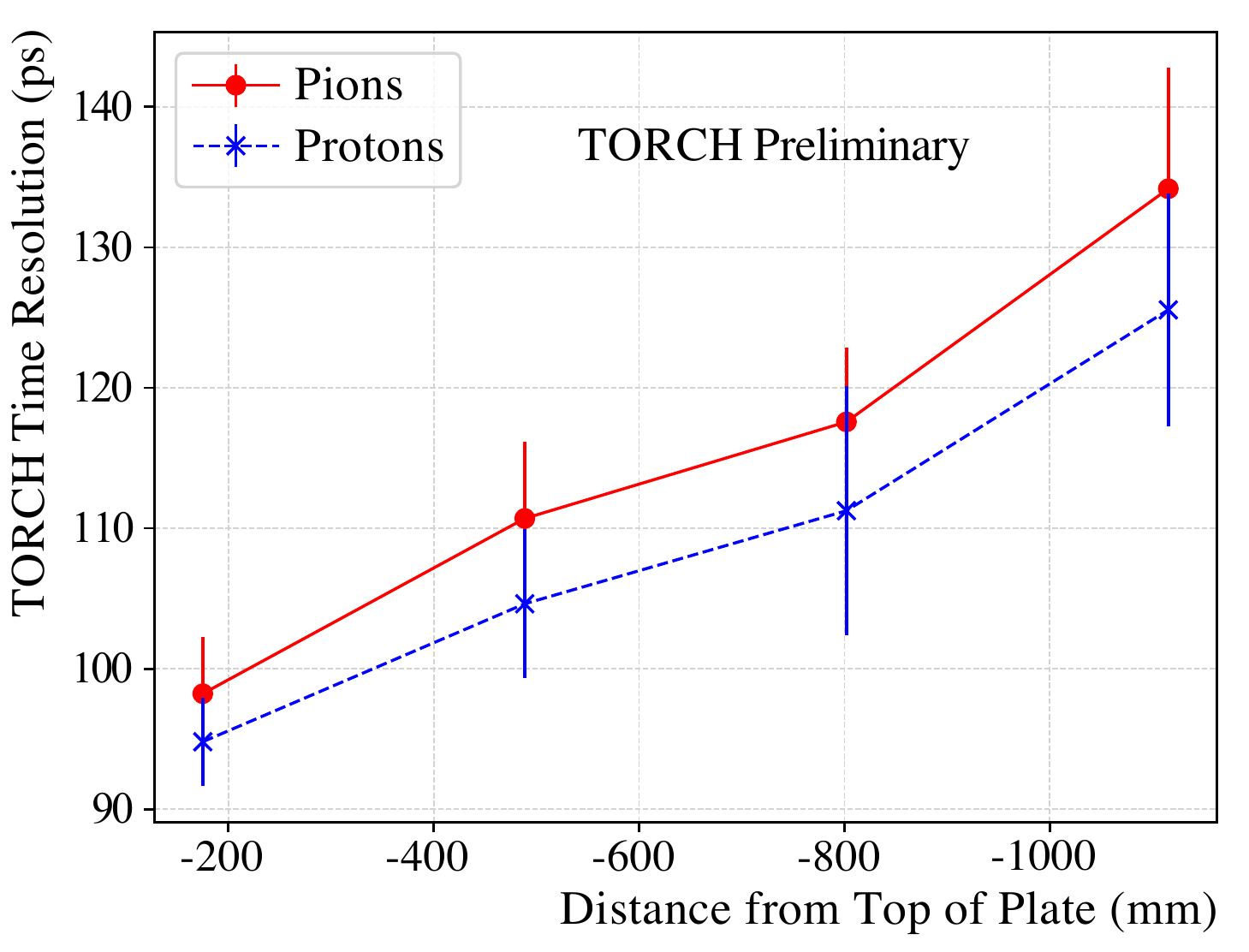}
\caption{
The dependence of the TORCH single-photon timing resolution as a function of the vertical position of the beam incident on the radiator plate.
The beam is aligned $5\,\rm{mm}$ from the side edge of the quartz, directly below the MCPs. The quoted resolution is averaged over each MCP-PMT column, and the error is the standard deviation of the measurements.
}
\label{fig:Time_Resolutions}
\end{figure}

Measurements of the timing resolutions are made with the beam incident $5\,\rm{mm}$ from the side edge of the quartz plate, directly below the MCP-PMTs.
This side is chosen over the horizontal midpoint to aid in the separation of different orders of reflection.
Currently only the more efficient MCP-PMT is used in the analysis, and the resolutions are determined separately for incident protons and pions, and also for each MCP-PMT column. 
\fig{fig:Time_Resolutions} shows the average, and standard deviation, of the resolution measurements over the MCP-PMT columns as a function of vertical displacement, for protons and pions.
The best achieved time resolution is $89.1\pm1.3\,\rm{ps}$. The resolution is marginally better for protons than pions, attributed to the proton sample being slightly purer than the pion sample. A degradation in time resolution is seen with increasing photon path length in the quartz, and this effect is under investigation.

Whilst a time-walk correction is already incorporated, the main factor currently limiting the resolution is the calibration for the time-over-threshold response of the NINO chip which yet needs to be introduced. 
Work is underway to incorporate a charge to signal-width calibration, providing a correction not only to the photon time of arrival but also the spatial positioning.

%
\section{Photon Counting}
\label{sec:Photon_Counting}
To investigate the photon counting efficiency of Proto-TORCH, the number of photons observed in data is compared to simulation. Optical processes are modelled by Geant4 \cite{Allison_2016}, while custom libraries are used to model the detector response and readout electronics. Both MCP-PMTs are used in the counting measurement, with the quantum efficiencies displayed in \fig{fig:Hitmap}b as input.

\begin{figure}[ht]
\centering
\includegraphics[width=0.95\linewidth]{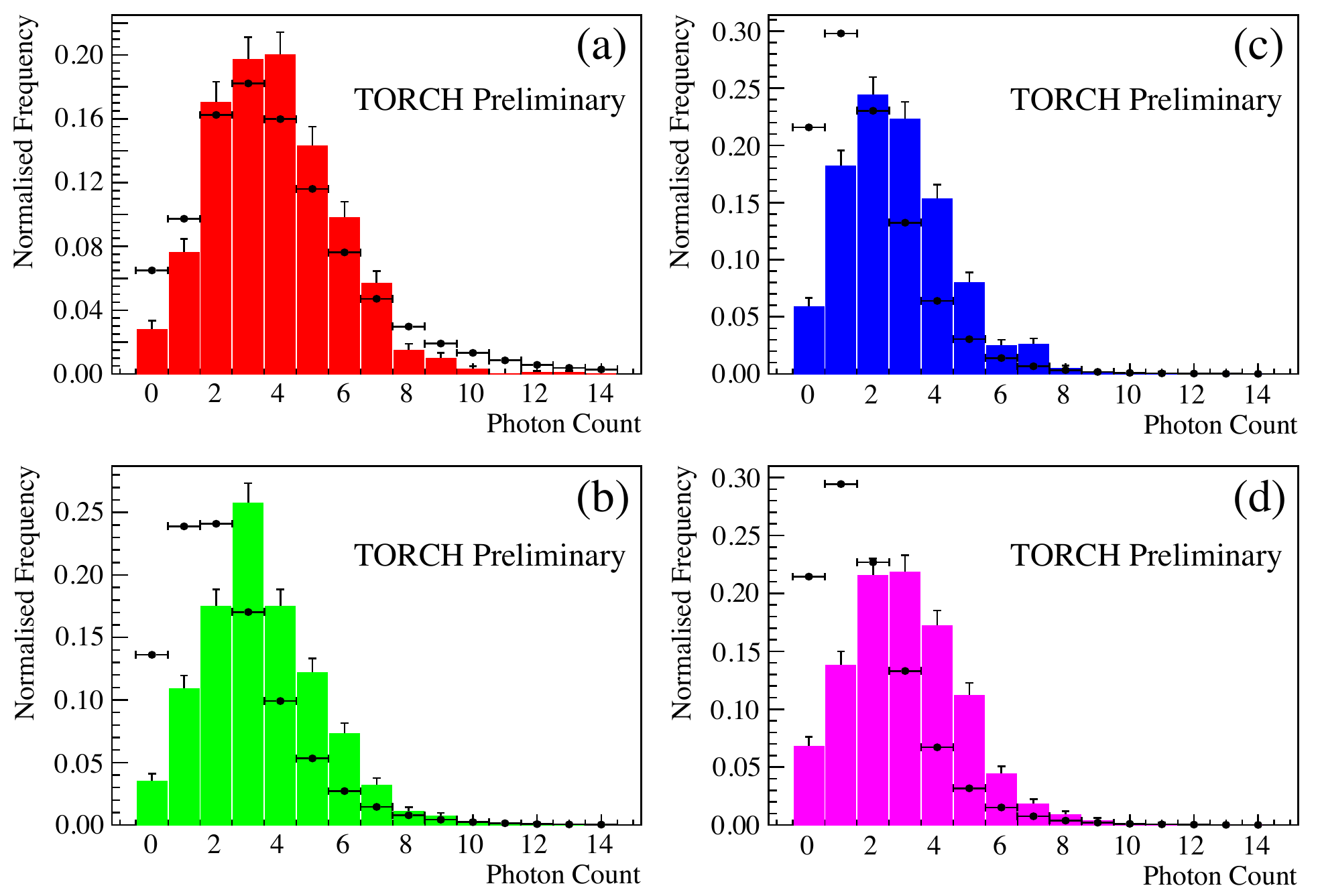}
\caption{
A comparison of the number of photons seen in data (points) and simulation (histograms) for different vertical beam positions on the radiator plate. The beam is aligned $5\,\rm{mm}$ from the side edge below the MCPs, with vertical displacements matching those in \fig{fig:Time_Resolutions}: (a) $-175\,\rm{mm}$, (b) $-488\,\rm{mm}$, (c) $-801\,\rm{mm}$, and (d) $-1115\,\rm{mm}$.
}
\label{fig:Photon_Counting}
\end{figure}

\fig{fig:Photon_Counting} shows the number of photons per incident particle in data for different vertical positions on the radiator plate compared to simulation. For beam position (a), located just below the MCP-PMTs, good agreement between data and simulation is observed. Moving down the plate, approximately $65\,\%$ more photons are seen in simulation than data. This indicates a loss of photons, possibly due to local surface scattering somewhere below the top point, and this effect is currently under study.

A geometrical factor of $\sim\,$6 increase in the number of photons is expected when scaling up the 2 MCP-PMTs, tested here, to a full module of 11. 
The final tubes can also be expected to have significantly better quantum efficiency on average than the tubes used in the test beam, bringing the photon yield towards the desired 30 photons per event.

%
\section{PID Performance}
\label{sec:PID_Performance}
The expected PID performance of the final TORCH detector in the LHCb experiment has been simulated. A $5 \times 6 \,\rm{m^{2}}$ area of quartz radiator is assumed, divided into 18 modules, each read out with 11 MCP-PMTs. At the present time the simulation is stand-alone, with events (and background) from the full LHCb simulation \cite{Clemencic_2011} used as input to model the behaviour.

A PID algorithm has been developed based on the delta log likelihood approach employed for the LHCb RICH detectors \cite{Forty_1999}. For each track incident on the radiator plate, a pattern of MCP-PMT hits is generated for each mass hypothesis. The detected hits are then compared to these patterns, and the combination of hypotheses which gives the greatest improvement in log-likelihood ($\Delta$LL) is chosen as the preferred set of particle types.

\begin{figure}[ht]
\centering
\includegraphics[width=0.85\linewidth]{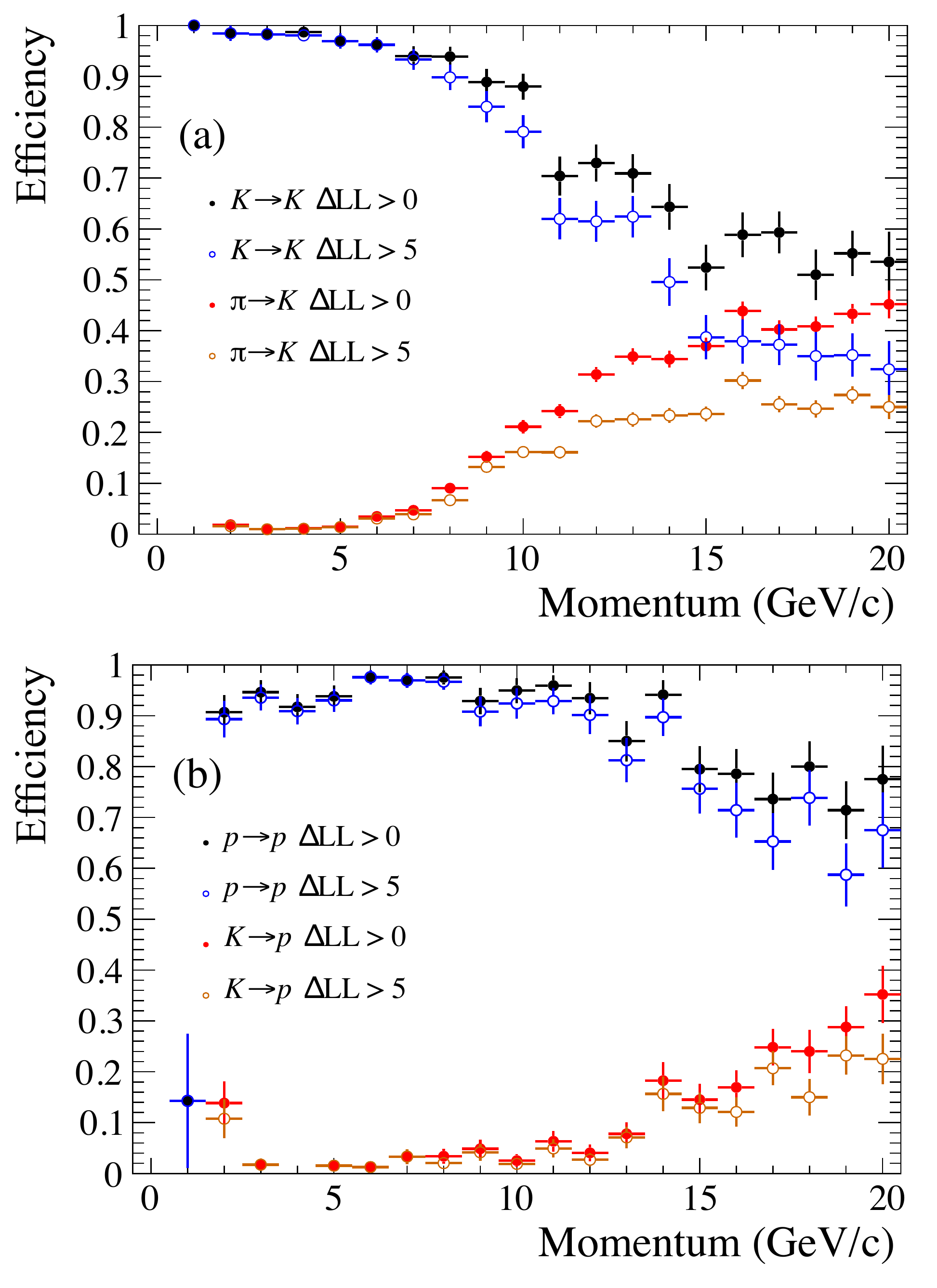}
\caption{
The PID efficiency of the full TORCH detector for (a) kaon-pion separation, (b) proton-kaon separation, determined from stand-alone simulation at an effective luminosity of $2\times10^{33}\,\rm{cm^{-2}s^{-1}}$. Also shown are corresponding misidentification fractions.
}
\label{fig:TORCH_PID}
\end{figure}

Pion-kaon and kaon-proton separation efficiencies and misidentification fractions have been investigated. Events generated with at least one $B$ meson at a luminosity of $\mathcal{L} = 2\times10^{33}\,\rm{cm^{-2}s^{-1}}$, including pile-up interactions, were used. This corresponds to the conditions expected at Run 3 of the LHC. TORCH has excellent separation power in the momentum range of interest ($2-10\,\rm{GeV/c}$) for all particle types, demonstrated in \fig{fig:TORCH_PID}. For the pion-kaon case, the efficiency is excellent up to $10\,\rm{GeV/c}$, beyond which some discrimination power remains; for the kaon-proton case the separating power extends up to $15\,\rm{GeV/c}$ and beyond.

Work is currently ongoing to simulate TORCH in LHCb up to luminosities of $\mathcal{L} = 2\times10^{34}\,\rm{cm^{-2}s^{-1}}$ (i.e. in Upgrade II conditions \cite{LHCb_Collab_2018}), as well as to implement TORCH into the full LHCb Monte Carlo framework.

\section{Conclusions}
\label{sec:Conclusions}
A half-sized TORCH module with a $660 \times 1250 \times 10\,\rm{mm^3}$ radiator plate has been tested in an $8\,\rm{GeV/c}$ mixed proton-pion beam at the CERN PS. The demonstrator employed customised $53 \times 53\,\rm{mm^{2}}$ MCP-PMTs with $8 \times 128$ pixel-equivalent granularity. Time resolutions have been measured which approach the desired $70\,\rm{ps}$ per photon, and improvements are anticipated with further calibrations of the electronics system. Photon yields are approximately $65\,\%$ of expectations, photon-path dependent, and which are the subject of further study. Simulations of TORCH in the LHCb experiment show that efficient pion-kaon and kaon-proton discrimination can be achieved up to $10\,\rm{GeV/c}$ and $15\,\rm{GeV/c}$, respectively.

\section*{Acknowledgements}
The support is acknowledged of the Science and Technology Research Council, UK, grant number ST/P002692/1, and of the European Research Council through an FP7 Advanced Grant (ERC-2011-AdG 299175-TORCH).

\newpage
\bibliography{Proceedings}

\begin{thebibliography}{1}
\expandafter\ifx\csname url\endcsname\relax
  \def\url#1{\texttt{#1}}\fi
\expandafter\ifx\csname urlprefix\endcsname\relax\def\urlprefix{URL }\fi
\expandafter\ifx\csname href\endcsname\relax
  \def\href#1#2{#2} \def\path#1{#1}\fi

\bibitem{Charles_2011}
M.~Charles, R.~Forty, {TORCH}: {T}ime of flight identification with {C}herenkov
  radiation, Nucl. Inst. \& Meth. in Phys. Res. A 639 (2011) 173 -- 176.
\newblock \href {http://arxiv.org/abs/1009.3793} {\path{arXiv:1009.3793}},
  \href {http://dx.doi.org/10.1016/j.nima.2010.09.021}
  {\path{doi:10.1016/j.nima.2010.09.021}}.

\bibitem{Brook_2018}
N.~Brook, et~al., Testbeam studies of a {TORCH} prototype detector, Nucl. Inst.
  \& Meth. in Phys. Res. A 908 (2018) 256 -- 268.
\newblock \href {http://arxiv.org/abs/1805.04849} {\path{arXiv:1805.04849}},
  \href {http://dx.doi.org/10.1016/j.nima.2018.07.023}
  {\path{doi:10.1016/j.nima.2018.07.023}}.

\bibitem{Conneely_2015}
T.~Conneely, et~al., The {TORCH} {PMT}: a close packing, multi-anode, long life
  {MCP-PMT} for {C}herenkov applications, Journal of Instrumentation 10 (2015)
  C05003.
\newblock \href {http://dx.doi.org/10.1088/1748-0221/10/05/C05003}
  {\path{doi:10.1088/1748-0221/10/05/C05003}}.

\bibitem{Akindinov_2004}
A.~Akindinov, et~al., Design aspects and prototype test of a very precise {TDC}
  system implemented for the {M}ultigap {RPC} of the {ALICE-TOF}, Nucl. Inst.
  \& Meth. in Phys. Res. A 533 (2004) 178 -- 182, proceedings of the Seventh
  International Workshop on Resistive Plate Chambers and Related Detectors.
\newblock \href {http://dx.doi.org/10.1016/j.nima.2004.07.023}
  {\path{doi:10.1016/j.nima.2004.07.023}}.

\bibitem{Anghinolfi_2004}
F.~Anghinolfi, et~al., {NINO}: an ultra-fast and low-power front-end
  amplifier/discriminator {ASIC} designed for the multigap resistive plate
  chamber, Nucl. Inst. \& Meth. in Phys. Res. A 533 (2004) 183 -- 187.
\newblock \href {http://dx.doi.org/10.1016/j.nima.2004.07.024}
  {\path{doi:10.1016/j.nima.2004.07.024}}.

\bibitem{Allison_2016}
J.~Allison, et~al., Recent developments in {G}eant4, Nucl. Inst. \& Meth. in
  Phys. Res. A 835 (2016) 186 -- 225.
\newblock \href {http://dx.doi.org/10.1016/j.nima.2016.06.125}
  {\path{doi:10.1016/j.nima.2016.06.125}}.

\bibitem{Clemencic_2011}
M.~Clemencic, et~al., The {LHCb} {S}imulation {A}pplication, {G}auss: {D}esign,
  {E}volution and {E}xperience, Journal of Physics: Conference Series 331
  (2011) 032023.
\newblock \href {http://dx.doi.org/10.1088/1742-6596/331/3/032023}
  {\path{doi:10.1088/1742-6596/331/3/032023}}.

\bibitem{Forty_1999}
R.~Forty, {RICH} pattern recognition for {LHCb}, Nucl. Inst. \& Meth. in Phys.
  Res. A 433 (1999) 257 -- 261.
\newblock \href {http://dx.doi.org/10.1016/S0168-9002(99)00310-1}
  {\path{doi:10.1016/S0168-9002(99)00310-1}}.

\bibitem{LHCb_Collab_2018}
{The {LHCb} {C}ollaboration, LHCb Experiment},
  \href{https://cds.cern.ch/record/2320509}{{Physics case for an {LHCb}
  {U}pgrade {II}}}, Tech. Rep. LHCb-PUB-2018-009. CERN-LHCb-PUB-2018-009, CERN,
  Geneva (May 2018).
\newblock \href {http://arxiv.org/abs/1808.08865} {\path{arXiv:1808.08865}}.
\newline\urlprefix\url{https://cds.cern.ch/record/2320509}

\end{thebibliography}

\newpage

\end{document}